\documentclass[showpacs,aps,prd,nofootinbib,floatfix,amsmath,amssymb]{revtex4}
\usepackage{graphicx}
\usepackage{physics}
\begin{document}

\makeatletter
\newbox\slashbox \setbox\slashbox=\hbox{$/$}
\newbox\Slashbox \setbox\Slashbox=\hbox{\large$/$}
\def\pFMslash#1{\setbox\@tempboxa=\hbox{$#1$}
  \@tempdima=0.5\wd\slashbox \advance\@tempdima 0.5\wd\@tempboxa
  \copy\slashbox \kern-\@tempdima \box\@tempboxa}
\def\pFMSlash#1{\setbox\@tempboxa=\hbox{$#1$}
  \@tempdima=0.5\wd\Slashbox \advance\@tempdima 0.5\wd\@tempboxa
  \copy\Slashbox \kern-\@tempdima \box\@tempboxa}
\def\FMslash{\protect\pFMslash}
\def\FMSlash{\protect\pFMSlash}
\def\miss#1{\ifmmode{/\mkern-11mu #1}\else{${/\mkern-11mu #1}$}\fi}
\makeatother

\title{Batalin-Vilkovisky description of Yang-Mills theories with universal extra dimensions}
\author{J. A. Ahuatzi-Avenda\~no, A. J. Correa-Le\' on, M. Huerta-Leal, H. Novales-S\' anchez, A. Sierra-Mart\'\i nez, and J. J. Toscano}
\address{Facultad de Ciencias F\'{\i}sico Matem\'aticas,
Benem\'erita Universidad Aut\'onoma de Puebla, Apartado Postal
1152, Puebla, Puebla, M\'exico. }
\begin{abstract}
The Batalin-Vilkovisky (BV) framework is a powerful technique for quantizing a wide range of gauge systems. This formalism employs fields and antifields—introducing a symplectic structure through the antibracket—to implement BRST symmetry, which captures the essence of gauge invariance. Within this context, we study the symmetry structure of a pure Yang-Mills theory with universal extra dimensions. We construct a BRST-invariant extended action for the \((4+n)\)-dimensional theory depending on fields and antifields, which constitutes a proper solution to the master equation. The higher-dimensional spacetime and gauge symmetries are then hidden within their four-dimensional counterparts via canonical transformations. Gauge invariances are fixed by combining gauge-fixing procedures for both the standard and Kaluza-Klein (KK) fields. This procedure is implemented covariantly in the adjoint representation of the gauge group. While the standard gauge fields are fixed using the Background Field Method (BFM), the KK gauge excitations are fixed covariantly, as they transform as matter fields. Finally, the manifest \(SU(N)\) gauge symmetry of the quantized theory is emphasized.
\end{abstract}

\maketitle

\section{Introduction}
\label{I}The hypothesis of additional spatial dimensions dates back over a century to the pioneering work of G. Nordström~\cite{Nordstrom}, T. Kaluza~\cite{Kaluza}, and O. Klein~\cite{Klein}, who originally attempted to unify electromagnetism with gravitation by extending the spacetime geometry. This geometric framework gained renewed impetus with the advent of string theory~\cite{Veneziano,Nielsen,KoNi1,KoNi2,Nambu,Susskind1,Susskind2}, whose original bosonic formulation required 26 spacetime dimensions to ensure quantum unitarity~\cite{Lovelace}. The subsequent incorporation of supersymmetry led to superstring theory~\cite{Ramond,WeZu,SchSch}, which successfully reduced the critical spacetime dimensionality from 26 to 10~\cite{Schwarz} while providing a consistent framework for quantum gravity. Further conceptual developments culminating in \(M\)-theory~\cite{Witten} revealed that the five known superstring formulations are merely distinct perturbative limits of a single, unifying fundamental theory naturally set in 11 spacetime dimensions. Consequently, extra dimensions evolved from a clever mathematical device into an indispensable structural requirement for modern unified theories.\\

Interest in the phenomenological implications of extra dimensions intensified following the proposals by Antoniadis, Arkani-Hamed, Dimopoulos, and Dvali~\cite{A,ADD,AADD}, who argued that relatively large extra dimensions could manifest at the TeV scale. Although no evidence of physics beyond the Standard Model (SM) has emerged at the LHC, new-physics effects remain plausible at future high-energy and high-luminosity frontiers. To probe these boundaries, the particle physics community is actively developing next-generation machines. These include precision \(e^{+}e^{-}\) linear colliders such as the International Linear Collider (ILC)~\cite{ILC1,ILC2} (upgradable to 1 TeV) and the Compact Linear Collider (CLIC)~\cite{CLIC1,CLIC2,CLIC3} (reaching up to 3 TeV), as well as circular proposals like the Circular Electron-Positron Collider (CEPC)~\cite{CEPC1,CEPC2} and its high-energy hadron phase, the Super Proton Proton Collider (SPPC)~\cite{JT,SPPC1,SPPC2}, designed to reach up to 125 TeV. Such future facilities will perform precision measurements crucial for detecting heavy physics signals via virtual effects on low-energy observables. Within this context, if universal extra dimensions exist, a robust effective field theory framework\footnote{Although such a framework contains interactions that are non-renormalizable in the Dyson sense, it remains fully predictive at accessible energy scales as a low-energy effective theory.} becomes essential to systematically analyze their impact on these observables.\\

In this paper, we discuss the theoretical foundations of the classical and quantum structures of field theories with compact extra dimensions. Given that the most compelling yet intricate aspects involve gauge invariance, we focus our discussion on pure Yang-Mills theory (without matter fields), adopting the well-known Universal Extra Dimensions (UED) approach~\cite{ACD1}. Building upon a previous study~\cite{OPRM} where we examined the gauge structure, degrees of freedom, and mass spectrum of the effective theory, the present work extends that analysis by implementing the Batalin-Vilkovisky (BV) or field-antifield formalism~\cite{BV1,BV2}. Within this framework, BRST symmetry—the cornerstone of gauge invariance—is manifest at the classical level, providing a robust and elegant basis for quantization.\\

The quantization of gauge systems is notoriously non-trivial due to the indefiniteness of the path integral measure. The BV formalism addresses this issue by extending the configuration space, introducing an antifield $\Phi^*_A$ for each field $\Phi^A$, and defining an extended action $S$. Within this extended space, a symplectic structure known as the antibracket, denoted as $(X,Y)$, is established, which resembles graded Poisson brackets~\cite{GPB}. The resulting master equation, $(S,S)=0$, ensures BRST invariance, while the subsequent gauge-fixing procedure is implemented via an appropriate gauge-fixing fermion $\Psi$~\cite{PR,DA}.\\

Our starting point is a theory with $n$ extra dimensions valid at energy scales $\Lambda \gg R^{-1}$, where $R$ denotes the size of the compact dimensions. At these scales, the compact dimensions appear effectively infinite, and the theory is governed by the extended groups $ISO(1,3+n)$ and $SU(N, {\cal M}^{4+n})$\footnote{We denote the gauge group $SU(N)$ in terms of its base manifold to clarify that the additional gauge fields result from higher dimensionality, rather than from enlarging the Lie algebra.}. The transition to the low-energy regime involves hiding these symmetries within the four-dimensional $ISO(1,3)$ and $SU(N,{\cal M}^4)$ groups. This dimensional reduction is achieved through a canonical transformation that maps high-energy covariant objects onto their low-energy counterparts~\cite{OPRM}. We implement this map within the field-antifield formalism to ensure that the antibracket structure remains strictly invariant. While preliminary results have been reported in specific contexts~\cite{PRDYM, PRDQED, JPGHIGGS, JPGS, OPT1, OPT2, OPT3}, a comprehensive treatment of the underlying structural basis remains absent. This paper fills that gap by providing a unified classical and quantum description; to the best of our knowledge, this is the first time the field-antifield formalism is applied to systematically implement the symmetry reduction of such extra-dimensional theories.\\

At energy scales of order $\Lambda$, the fundamental degrees of freedom are the gauge fields ${\cal A}^a_M(x,\bar x)$ of the extended gauge group $SU(N,{\cal M}^{4+n})$. Here, $M=\mu, \bar \mu$ is an index of the extended Lorentz group $SO(1,3+n)$, where $\mu=0,1,2,3$ and $\bar \mu=5,\cdots,4+n$ denote the indices of the standard Lorentz group $SO(1,3)$ and the orthogonal group $SO(n)$ defined on the Minkowski ${\cal M}^4$ and Euclidean ${\cal N}^n$ manifolds, respectively; their arguments are points on these respective manifolds, namely $x\in {\cal M}^4$ and $\bar x \in {\cal N}^n$. Within the BV framework, the standard configuration space is extended by treating the gauge parameters $\alpha^a(x,\bar x)$ as genuine degrees of freedom necessary for quantization. This is achieved by introducing an anticommuting (fermionic) ghost field ${\cal C}^a (x,\bar x)$ for each parameter $\alpha^a(x,\bar x)$. Each field $\Phi^A=\{{\cal A}^{a M}(x,\bar x), {\cal C}^a(x,\bar x)\}$ is paired with a corresponding antifield $\Phi^*_A(x,\bar x)=\{ {\cal A}^*_{a M}(x,\bar x),{\cal C}^*_a(x,\bar x)\}$, which transforms under the same representations of the extended groups but carries opposite statistics. The minimal proper solution $S_m[\Phi^A, \Phi^*_A]$ of the master equation remains degenerate and cannot be directly used in the path integral. To lift this degeneracy, a gauge-fixing procedure is introduced by incorporating trivial pairs $(\bar{\cal{C}}^a, {\cal }B^a)$—consisting of fermionic antighosts and bosonic auxiliary fields—along with their respective antifields. These antighosts are essential for constructing the gauge-fixing fermion $\Psi$, a functional that must carry a ghost number ($gh$) of $-1$. To satisfy this requirement, the antighosts are assigned $gh(\bar{{\cal C}}^a)=-1$, contrasting with the ghosts where $gh(C^a)=1$. Ultimately, eliminating the antifields through the identification $\Phi^*_A=\delta \Psi/\delta \Phi^A$ breaks the underlying gauge degeneracy of the theory.\\

To transition from the high-energy theory to the low-energy effective description, we implement a canonical transformation—in the sense of the antibracket—that maps the high-energy fields $\Phi^A(x,\bar x)$ and antifields $\Phi^*_A(x,\bar x)$ into covariant objects of the standard groups $ISO(1,3)\times SU(N,{\cal M}^4)$. As shown below, the low-energy degrees of freedom manifest as an infinite tower of Kaluza-Klein (KK) excitations of fields $\{\Phi^{(\underline{0})A}(x), \Phi^{(\underline{k})A}(x)\}$ and antifields $\{\Phi^*_{(\underline{0})A}(x), \Phi^*_{(\underline{k})A}(x)\}$, where $(\underline{k})$ denotes the Fourier indices and $(\underline{0})$ represents the zero modes. Following Ref.~\cite{OPRM}, the higher-dimensional gauge field ${\cal A}^{aM}(x,\bar x)$ decomposes into the zero mode $A^{(\underline{0}) a\mu}(x)$ and a set of KK excitations. The latter includes massive gauge fields $A^{(\underline{k})a\mu}(x)$, associated pseudo-Goldstone bosons $A^{(\underline{k})a}_G(x)$, and $n-1$ physical scalars $A^{(\underline{k})a}_{\bar n}(x)$, all transforming under the adjoint representation of $SU(N,{\cal M}^4)$.\\

To preserve manifest gauge invariance under the low-energy $SU(N,{\cal M}^4)$ group within the effective action, we employ the Background Field Method (BFM)~\cite{BFM}. This technique ensures that the action remains explicitly invariant under background gauge transformations by splitting each field into classical and quantum components, where only the latter are integrated over in the path integral. Specifically, the zero-mode gauge field is decomposed as $A^{(\underline{0})a\mu}(x)\to \bar{A}^{(\underline{0})a\mu}(x)+Q^{(\underline{0})a\mu}(x)$, where $\bar{A}^{(\underline{0})a\mu}(x)$ and $Q^{(\underline{0})a\mu}(x)$ represent the background and quantum components, respectively. While all fields—including the KK excitations—could in principle undergo a similar split, we restrict this decomposition to the standard fields. Given that our main interest concerns the one-loop contributions of KK excitations to standard observables, we do not introduce background components for the KK excitations or the ghost fields. This choice is justified because, as demonstrated in the following sections, the gauge-fixing for the KK excitations $A^{(\underline{k})a \mu}(x)$ can be performed in a manner that is naturally covariant under the $SU(N,{\cal M}^4)$ group.\\

The rest of the paper is organized as follows. In Sec.~\ref{FAF}, the classical BRST structure of Yang-Mills theory in $(4+n)$-dimensions is examined within the field-antifield formalism. We outline the conceptual and technical ingredients required to derive the extended four-dimensional theory and analyze its BRST structure in the context of the BFM. Sec.~\ref{LET} addresses the reduction of the $(4+n)$-dimensional theory to the 4-dimensional one within the context of the field-antifield formalism. Finally, Sec.~\ref{C} presents our conclusions.

\section{The Field-Antifield Formulation}
\label{FAF}
In accordance with our discussion presented in the Introduction, our starting point is an effective gauge field theory governed by the extended group ${\rm ISO}(1,3+n) \times {\rm SU}(N,{\mathcal M}^{d})$, and whose gauge parameters are defined all over the spacetime $\mathcal{M}^{d}=\mathcal{M}^{4}\times\mathcal{N}^{n} $. Since the theory is not renormalizable in the Dyson's sense, the corresponding action consists of an infinite series of Lorentz and gauge invariant terms of increasing canonical dimension, that is,
\begin{equation}
\label{ea}
S_{\rm eff}[{\cal A}^a_M]=\int d^{\,4}x\,d^{\,n}\bar{x} \,\,\left[{\cal L}^d_{YM}+\sum_{\textbf{d}} \frac{\lambda_{\textbf{d}}}{\Lambda^{\textbf{d}}}\,{\cal L}^{(\textbf{d}+d)}\left({\cal F}^a_{MN},{\cal D}^{ab}_A {\cal F}^b_{MN}\right)\right]\, ,
\end{equation}
where
\begin{eqnarray}
\label{LGD}
 {\cal L}^d_{YM}=-\frac{1}{4}{\cal F}^a_{MN}(x,\bar x){\cal F}^{MN}_a(x,\bar x)\, ,
\end{eqnarray}
with
\begin{equation}
\label{YMC}
{\cal F}^a_{MN}(x,\bar x)=\partial_M {\cal A}^a_N(x,\bar x)-\partial_N {\cal A}^a_M(x,\bar x)+g_df^{abc}{\cal A}^b_M(x,\bar x){\cal A}^c_N(x,\bar x)\, .
\end{equation}
In this expression, $g_{d}$ and $f^{abc}$ are the coupling constant and the structure constants of the ${\rm SU}(N,{\cal M}^d)$ group. We will use a metric with signature $g=diag(1,-1,-1,\cdots)$. Under the extended gauge group ${\rm SU}(N,{\cal M}^{d})$, the connection and curvature components transform, respectively, as
\begin{subequations}
\begin{align}
\label{GT}
\delta {\cal A}^a_M(x,\bar x)&={\cal D}^{ab}_M \alpha^b(x,\bar x)\, , \\
\label{CT}
\delta {\cal F}^a_{MN}(x,\bar x)&=g_{d}f^{abc} {\cal F}^b_{MN}(x,\bar x)\alpha^c(x,\bar x)\, ,
\end{align}
\end{subequations}
where $\alpha^a(x,\bar x)$ are the gauge parameters and ${\cal D}^{ab}_M=\delta^{ab}\partial_M-g_{d}f^{abc}{\cal A}^c_M(x,\bar x)$ is the covariant derivative in the adjoint representation of the group. In Eq.~(\ref{ea}), ${\cal L}^{(\textbf{d}+d)}$ represents gauge- and Lorentz-invariant interactions of canonical dimension greater than $d=4+n$, formulated from curvatures ${\cal F}^a_{MN}$ and its covariant derivatives ${\cal D}^{ab}_{M} {\cal F}^{b}_{MN}$, multiplied by unknown coupling constants $ \lambda_{\textbf{d}}/\Lambda^{\textbf{d}} $. The Lagrangian given by  Eq.~\eqref{LGD} corresponds to a straightforward extension of the functional structure of the well-known Yang-Mills Lagrangian from four dimensions to $ d $ dimensions. Terms of greater canonical dimension are suppressed by inverse powers of a fundamental scale $\Lambda$, which is assumed to be far above the explored energies. In fact, according to the effective Lagrangian approach, this theory is valid only for energies below $\Lambda$. As already commented in the Introduction, we assume that the size of the extra dimensions is so large compared with these energies that it actually can be considered as infinite. This justifies the ${\rm ISO}(1,3+n)$ description of the effective theory~\eqref{ea}. Note that the Lagrangian of Eq.~\eqref{LGD} does not depend on the $\Lambda$ scale, although it does depend on the dimensionful coupling constant $g_{d}$, which must be rescaled to obtain the correct dimensionless Yang-Mills coupling. As it is shown in Ref.~\cite{OPRM}, this rescaling of the coupling constant is naturally given by the compactification scale $R^{-1}$, since only then is it possible to make contact with the usual theory. As argued in~\cite{OPRM}, the dominant contributions arise from the $d$-dimensional version of the standard theory given in Eq.~(\ref{LGD}); thus, we shall henceforth focus exclusively on this part. However, the terms of dimension greater than $d$, ${\cal L}^{(\textbf{d}+d)}$, play a central role in the renormalization of the theory within the framework of effective field theories.\\

We will focus on the $d$-dimensional version of the Yang-Mills theory given by Lagrangian~(\ref{LGD}). The corresponding action will be denoted by
\begin{equation}
\label{CA}
S_0[{\cal A}^a_M]=\int d^4x\,d^n\bar{x} \left[-\frac{1}{4}{\cal F}^a_{MN}(x,\bar{x}){\cal F}_a^{M N}(x,\bar{x})\right]\, .
\end{equation}
As anticipated in the introduction, within the BV formalism, the ghost fields are recognized as genuine degrees of freedom because they are necessary to quantize the theory. So the original configuration space is extended to include a ghost field per each gauge parameter. It is assumed that the ghost-field statistics is opposite to the statistics of the gauge parameters, so in our case, the ghost fields, which will be denoted by ${\cal C}^a(x,\bar{x})$, are anticommuting or fermion fields. Consequently, our fields $\Phi^A$ are ${\cal A}^{aM}(x,\bar{x})$ and ${\cal C}^a(x,\bar{x})$, which are the gauge and ghost fields associated with the ${\rm SU} (N,{\cal M}^d)$ gauge group. An additive conserved charge, called the ghost number and denoted by $gh$, is assigned to each field. Classical fields, in our case the ${\cal A}^{aM}(x,\bar{x})$ gauge fields, have ghost number $0$, while the ghost number of the ghost fields is $1$. This configuration space is further extended by introducing an antifield $\Phi^*_A (x,\bar{x})$ per each field $\Phi^A(x,\bar{x})$, that is, we have antifields $\Phi^*_A (x,\bar{x})=\{{\cal A}^*_{aM}(x,\bar{x}), {\cal C}^*_a(x,\bar{x})\}$. The statistics of any antifield is opposite to that of its associated field, whereas the corresponding ghost numbers are related as ${\rm gh}[\Phi^*_A(x,\bar{x})]=-{\rm gh}[\Phi^A(x,\bar{x})]-1$. Although antifields are only a mathematical tool of formalism and do not represent physical degrees of freedom, their introduction provides us with an unambiguous criterion to introduce a gauge-fixing  procedure, which is necessary to quantize the theory. The antifields do not contribute to the fundamental path integral, since the gauge-fixing procedure is precisely defined to eliminate them.\\

In the space of fields and antifields, the antibracket is defined through left and right derivatives as follows:

\begin{equation}
(X,Y)=\frac{\delta_R X}{\delta \Phi^A}\frac{\delta_L Y}{\delta \Phi^*_A}-\frac{\delta_R X}{\delta \Phi^*_A}\frac{\delta_L Y}{\delta \Phi^A}\, ,
\end{equation}
where $X$ and $Y$ are functionals on fields and antifields. The antibracket has properties~\cite{PR} that are similar to those of the graded Poisson's bracket~\cite{GPB}. From this definition, the fundamental antibracket
\begin{equation}
\label{FAB}
(\Phi^A,\Phi^*_B)=\delta^A_B\,
\end{equation}
arises as a particular case, showing us that the fields and antifields constitute canonical pairs, just as the usual phase space is defined by canonical pairs. In this formulation, the classical action is extended to include, besides the ghost fields, the antifields. This extended action, which is denoted by $S[\Phi, \Phi^*]$, is a real bosonic functional with ghost number $0$, which satisfies the classical master equation
\begin{equation}
\label{ME}
(S,S)=0\, .
\end{equation}
The solution to the master equation plays a double role, since, on the one hand, it is the generating functional of the gauge structures and, on the other hand, it constitutes the starting point to quantize the theory. Not all possible solutions of the master equation are physically acceptable, since certain boundary conditions must be satisfied~\cite{PR}. An essential requirement is the classical limit $S[\Phi,\Phi^*]|_{\Phi^*=0}=S_0[{\cal A}^{aM}]$. A solution that fulfills this is called a proper solution, which can be expressed as power series in the antifields.\\

 At the classical level, the BRST symmetry involves both fields and antifields, and describes a theory that contains degeneration, that is, a theory with gauge invariances. This means that a gauge-fixing procedure must be introduced to define the fundamental path integral, in the implementation of which antifields play a central role. It turns out that, after introducing the gauge-fixing procedure, the nondegenerate action is invariant under a residual BRST symmetry, which is the quantum BRST symmetry~\cite{BRST}. First, we briefly discuss the main features of classical BRST symmetry and its importance in the quantization  of gauge systems. The generator of the classical BRST symmetry is the proper solution $S$ through the antibracket, which is given by:
\begin{equation}
\label{BT}
\delta_B X=(X,S)\, ,
\end{equation}
where $X$ is an arbitrary functional on fields and antifields. Therefore, the master equation is nothing else than the statement that action $S$ is BRST invariant. The BRST transformations of the fields and antifields are given by
\begin{subequations}
\begin{align}
\label{bf}
\delta_B \Phi^A&=(\Phi^A,S)=\frac{\delta_L S}{\delta \Phi^*_A} \, ,\\
\label{baf}
\delta_B \Phi^*_A&=(\Phi^*_A,S)=-\frac{\delta_L S}{\delta \Phi^A} \, .
\end{align}
\end{subequations}
The proper solution for Yang-Mills theories is well known~\cite{PR}. Such a solution involves the gauge structures, which, in the case of Yang-Mills theories, correspond to the gauge generator, ${\cal D}^{abM}$, and the structure functions, given by the commutator of two gauge transformations, which in this case coincide with the structure constants of the group $SU(N)$, $f^{abc}$. So, in our case, such proper solution, also known as minimal solution, is given by
\begin{eqnarray}
\label{PS}
S_m[\Phi,\Phi^*]&=&\int d^4x\,d^n\bar{x} \Biggl \{ -\frac{1}{4}{\cal F}^a_{MN}(x,\bar{x}){\cal F}_a^{M N}(x,\bar{x})+{\cal A}^*_{aM}(x,\bar{x}){\cal D}^{abM}{\cal C}^b(x,\bar{x})\nonumber \\
&&+\frac{g_{d}}{2}f^{abc}{\cal C}^*_{a}(x,\bar{x}){\cal C}^b(x,\bar{x}) {\cal C}^c(x,\bar{x}) \Biggr \} \, .
\end{eqnarray}
From this expression, we can see that the extended action is the generating functional of the gauge structures, which follows from Eq.~(\ref{bf})
\begin{subequations}
\begin{align}
\label{bA}
\delta_B {\cal A}^{aM}(x,\bar x)&={\cal D}^{abM}{\cal C}^b(x,\bar{x})\, , \\
\label{bC}
\delta_B {\cal C}^a(x,\bar x)&=\frac{g_{d}}{2}f^{abc}{\cal C}^b(x,\bar{x}) {\cal C}^c(x,\bar{x})\, .
\end{align}
\end{subequations}
On the other hand, the BRST transformations of the antifields are given by:
\begin{subequations}
\begin{align}
\label{bAC}
\delta_B {\cal A}^*_{aM}(x,\bar x)&=-{\cal D}^{abN}{\cal F}_{NM}^b(x,\bar x)+g_{d}f^{abc}{\cal A}^*_{cM}(x,\bar x){\cal C}^b(x,\bar x)\, , \\
\delta_B {\cal C}^*_a(x,\bar x)&=-{\cal D}^{abM}{\cal A}^*_{b M}(x,\bar x)+g_{d}f^{abc}{\cal C}^*_c(x,\bar x){\cal C}^b(x,\bar x)\, .
\end{align}
\end{subequations}

\subsection{The gauge-Fixing Fermion $\Psi$}

The extended action $S_m$ cannot be quantized directly because it is gauge invariant. To define the fundamental path integral, one needs to introduce a gauge-fixing procedure to remove the degeneration caused by the gauge freedom from the theory. Since the antifields do not represent degrees of freedom, the path integral must be defined only in terms of the fields. Thus, the antifields must be removed from the theory before quantization. Note that the antifields cannot simply be equated to $0$, since this would lead us to the original theory, which is gauge invariant. However, one can remove the antifields in a nontrivial way and simultaneously remove the degeneration of the theory. In other words, the introduction of a mechanism to remove the  antifields is itself a gauge-fixing procedure. Following Batalin and Vilkovisky~\cite{BV1,BV2}, we introduce a functional $\Psi$, which depends only on the fields, such that
\begin{equation}
\label{psi}
\Phi^*_A=\frac{\delta \Psi[\Phi]}{\delta \Phi^A}.
\end{equation}
From this relation, we can establish the statistics and the ghost number that the functional $\Psi$ must have. Using the fact that the statistics and ghost number of the antifields are respectively given by $\epsilon(\Phi^*_A)=\epsilon(\Phi^A)+1 \, ({\rm mod} \, 2)$ and ${\rm gh}[\Phi^*_A]=-{\rm gh}[\Phi^A]-1$, we can see that $\Psi$ is a fermionic functional with ghost number equal to $-1$. The gauge-fixed action, which has no degeneration, is $S_\Psi=S[\Phi,\frac{\delta \Psi}{\delta \Phi^A}]$. As we have seen, to be consistent Eq.~(\ref{psi}) requires the functional $\Psi$ to have ghost number $-1$, which cannot happen since all of the fields of the theory have non-negative ghost numbers. To construct a $\Psi$ with the required ghost number, one introduces a trivial pair of fields $\{{\cal B}^a,\bar{{\cal C}}^a\}$ per each gauge field, with ghost numbers $0$ and $-1$, respectively. By definition, the statistics of ${\cal B}^a$ coincides with the statistics of the corresponding gauge parameter, whereas the statistics of $\bar{{\cal C}}^a$ is defined to be the opposite of the statistics of such a gauge parameter. In Yang-Mills theories, ${\cal B}^a$ and $\bar{{\cal C}}^a$ are even and odd fields, respectively. The bosonic fields ${\cal B}^a$ do not propagate, that is, they are auxiliary fields that can be removed from the theory through the equations of motion. On the other hand, the $\bar{{\cal C}}^a$ fermion fields are the Faddeev-Popov antighosts of the ghosts ${\cal C}^a$. The proper solution to the master equation is unique up to a canonical transformation and the addition of trivial pairs,
\begin{equation}
\label{trivial}
S_{\rm trivial}=\int d^4x d^n\bar{x}\, \bar{{\cal C}}^*_a(x,\bar{x}){\cal B}^a(x,\bar{x})\, ,
\end{equation}
where $\bar{{\cal C}}^*_a$ is the antifield of the $\bar{{\cal C}}^a$ field, which is bosonic and has ghost number $0$. Note that since neither ${\cal B}^*_a$ nor $\bar{{\cal C}}^a$ appears, the trivial pair contribution in $(S,S)$ is null. The proper non-minimal solution of the master equation is given by the sum of Eqs. (\ref{PS}) and (\ref{trivial}),
\begin{eqnarray}
\label{PSF}
S[\Phi,\Phi^*]&=&\int d^4x d^n\bar{x}\Biggl \{ -\frac{1}{4}{\cal F}^a_{MN}(x,\bar{x}){\cal F}_a^{M N}(x,\bar{x})+{\cal A}^*_{aM}(x,\bar{x}){\cal D}^{abM}{\cal C}^b(x,\bar{x})\nonumber \\
&&+\frac{g_{d}}{2}f^{abc}{\cal C}^*_{a}(x,\bar{x}){\cal C}^b(x,\bar{x}) {\cal C}^c(x,\bar{x})+\bar{{\cal C}}^*_a(x,\bar{x}){\cal B}^a(x,\bar{x})\Biggr \} \, ,
\end{eqnarray}
where to simplify the notation, the non-minimal solution will simply be denoted by $S$. From this action and equation ~(\ref{bf}) we see that the BRST transformation of the anti-ghost fields is
\begin{equation}
\label{bac}
\delta_B\bar{{\cal C}}^a(x,\bar x)={\cal B}^a(x,\bar x)\, ,
\end{equation}
so the auxiliary fields play the role of gauge structures in the context of the BRST symmetry. Note that $\delta_B \bar{\cal C}^*_a=0$.\\

We are now in a position to define the gauge-fixing fermion $\Psi$. Choosing a particular $\Psi$ is the same as choosing a gauge-fixing procedure. We choose a $\Psi$ given by
\begin{equation}
\label{FerFun}
\Psi[\Phi]=\int d^4x d^n\bar{x}\, \bar{{\cal C}}^a(x,\bar{x})\left[f^a+\frac{\xi}{2}{\cal B}^a(x,\bar{x})\right]\, ,
\end{equation}
where $f^a$ is a function on the fields $\Phi^A$ and its derivatives, whereas $\xi$ is the gauge-fixing parameter. Note that, like ${\cal B}^a$, $f^a$ must be bosonic, with ghost number $0$. This implies that $\Psi$ is fermionic and has ghost number -1, as required. If the gauge-fixing  functions $f^a$ are linear in the fields, they fulfill the basic purpose of defining the propagators of the gauge fields but without affecting the vertices of the theory. On the other hand, renormalization in the Dyson's sense allows defining gauge-fixing  functions depending quadratically on the gauge fields. These nonlinear gauge-fixing procedures not only allow us to define the propagators of the gauge fields but also non-trivially affect the gauge vertices of the theory. One important goal of this work is to introduce a gauge-fixing procedure that allows us to construct a fundamental path integral that preserves gauge invariance with respect the standard group ${\rm SU}(N,{\mathcal M}^{4})$. But before implementing an appropriate gauge-fixing procedure for this purpose, we must move from the high-energy description to the low-energy description.

\section{The Low-Energy theory}
\label{LET}
To preserve the essence of the theory, canonical transformations must be applied when transitioning from the high- to the low-energy regime. Consequently, we need to define canonical maps that allow us to descend towards low-energy regimes in accordance with the symmetry-hiding pattern that we have commented in the Introduction. As discussed in~\cite{OPRM}, this program can be started by introducing the canonical mappings directly into the gauge invariant action~(\ref{ea}) and thus go from the perspective ${\rm ISO}(1,3+n) \times {\rm SU}(N,{\mathcal M}^{d})$ to the perspective ${\rm ISO}(1,3) \times {\rm SU}(N,{\mathcal M}^{4})$. In this work, the mappings that allow us to make this transition will be implemented at the level of the extended action, which is invariant under classical BRST transformations. This will be the starting point to introduce a covariant gauge-fixing procedure based on the BFM and thus derive an effective theory with manifest ${\rm ISO}(1,3) \times {\rm SU}(N,{\mathcal M}^{4})$ symmetry.\\

The mapping details required to pass from the $d$-dimensional Yang-Mills theory (\ref{LGD}) to an effective 4-dimensional theory are provided in Ref.~\cite{OPRM}. Thus, we restrict ourselves here to a brief description of the fundamental steps within the field-antifield formalism. To hide the symmetry ${\rm ISO}(1,3+n) \times {\rm SU}(N,{\cal M}^d)$ into ${\rm ISO}(1,3) \times {\rm SU}(N,{\cal M}^4)$ means to hide the ${\rm ISO}(1,3+n)$ symmetry into ${\rm ISO}(1,3)$, since implicit to this is the passing from ${\rm SU}(N,{\cal M}^d)$ to ${\rm SU}(N,{\cal M}^4)$. We first map covariant objects of $ {\rm SO}(1,3+n) $ into covariant objects of its subgroups ${\rm SO}(1,3)$ and ${\rm SO}(n)$. The gauge fields ${\cal A}^a_M(x,\bar x)$, which are vector fields of $ {\rm SO}(1,3+n) $, can be seen as an $ {\rm SO}(1,3) $-vector field, with components $ \mathcal{A}^a_{\mu} $, and  $ n $ $ {\rm SO}(1,3) $-scalars denoted by $ \mathcal{A}^a_{\bar{\mu}} $; under the ${\rm SO}(n)$ group, the four components of $ \mathcal{A}^a_{\mu} $ are scalar fields, whereas $ \mathcal{A}^a_{\bar{\mu}} $ can be seen as a vector. Because the antifields $\Phi^*_A$ and the fields $\Phi^A$ transform under the same Lorentz group representations, these considerations extend naturally to the antifield sector,
\begin{subequations}
\begin{align}
{\rm SO}(1,3+n) &\mapsto  {\rm SO}(1,3)\times {\rm SO}(n) \, ,\nonumber  \\
\label{MF1}
{\cal A}^{aM}(x,\bar x)  &\mapsto  \{ {\cal A}^{a\mu} (x,\bar x), \, {\cal A}^{a{\bar \mu}} (x,\bar x)\}\, , \\
\label{MF2}
{\cal A}^*_{aM}(x,\bar x)  &\mapsto  \{ {\cal A}^*_{a\mu} (x,\bar x), \, {\cal A}^*_{a{\bar \mu}} (x,\bar x)\}\, .
\end{align}
\end{subequations}
The ghost fields, ${\cal C}^a(x, \bar x)$, the antighost fields, $\bar{{\cal C}}^a(x,\bar x)$, and the auxiliary fields, ${\cal B}^a(x,\bar x)$, as well as their corresponding antifields are Lorentz scalars and are, therefore, not affected by this transformation. This map in turn leads to the curvatures
\begin{equation}
\label{MC1}
{\cal F}^a_{MN}(x,\bar x) \mapsto \{ {\cal F}^a_{\mu \nu}(x,\bar x), \, {\cal F}^a_{\mu \bar \nu }(x,\bar x), \, {\cal F}^a_{\bar \mu \bar \nu}(x,\bar x)\} \, ,
\end{equation}
 with ${\cal F}^a_{\mu \nu}(x,\bar x)$, ${\cal F}^a_{\mu \bar \nu }(x,\bar x)$, and ${\cal F}^a_{\bar \mu \bar \nu}(x,\bar x)$ transforming as 2-form (0-form), 1-form (1-form), and 0-form (2-form) under the ${\rm SO}(1,3)({\rm SO}(n))$ group, respectively.\\

The maps (\ref{MF1}) and (\ref{MC1}) allow us to pass from the action given in Eq.~(\ref{CA}) to an action given by
\begin{eqnarray}
\label{A2}
S_0[{\cal A}^a_\mu,{\cal A}^a_{\bar \mu}]&=&-\frac{1}{4} \int d^4x\, d^n \bar x \Big[{\cal F}^a_{\mu \nu}(x,\bar x){\cal F}^{\mu \nu}_a(x,\bar x)+2{\cal F}^a_{\mu \bar \nu}(x,\bar x){\cal F}^{\mu \bar \nu}_a(x,\bar x)\nonumber \\
&&+{\cal F}^a_{\bar \mu \bar \nu}(x,\bar x){\cal F}^{\bar \mu \bar \nu}_a(x,\bar x)
\Big]\, ,
\end{eqnarray}
which is manifestly invariant under the ${\rm SO}(1,3)$ and ${\rm SO}(n)$ groups. Evidently, the map (\ref{MF1},\ref{MF2}) is canonical, as the fundamental antibracket $({\cal A}^{a M},{\cal A}^*_{bN})=\delta^a_b \delta^M_N$ is trivially mapped into the antibrackets $({\cal A}^{a \mu},{\cal A}^*_{b\nu})=\delta^a_b \delta^\mu_\nu$ and $({\cal A}^{a \bar \mu},{\cal A}^*_{b \bar \nu})=\delta^a_b \delta^{\bar \mu}_{\bar \nu}$. After these mappings, the gauge transformation (\ref{GT}) becomes
\begin{subequations}
\begin{align}
\label{LG1}
\delta {\cal A}^a_\mu(x,\bar x)&={\cal D}^{ab}_\mu \alpha^b(x,\bar x)\, , \\
\label{LG2}
\delta {\cal A}^a_{\bar \mu}(x,\bar x)&={\cal D}^{ab}_{\bar \mu} \alpha^b(x,\bar x)\, ,
\end{align}
\end{subequations}
while the BRST transformations (\ref{bA}) and (\ref{bAC}) take, respectively, the forms
\begin{subequations}
\begin{align}
\label{bA1}
\delta_B {\cal A}^{a\mu}(x,\bar x)&={\cal D}^{ab\mu}{\cal C}^b(x,\bar{x})\, , \\
\label{bA2}
\delta_B {\cal A}^{a\bar \mu}(x,\bar x)&={\cal D}^{ab\bar \mu}{\cal C}^b(x,\bar{x})\, ,
\end{align}
\end{subequations}

\begin{subequations}
\begin{align}
\label{bAC1}
\delta_B {\cal A}^*_{a\mu}(x,\bar x)&=-{\cal D}^{ab\nu}{\cal F}_{\nu \mu}^b(x,\bar x)-{\cal D}^{ab\bar \nu}{\cal F}_{\bar \nu \mu}^b(x,\bar x)+g_{d}f^{abc}{\cal A}^*_{c\mu}(x,\bar x){\cal C}^b(x,\bar x)\, , \\
\label{bAC2}
\delta_B {\cal A}^*_{a\bar \mu}(x,\bar x)&=-{\cal D}^{ab\nu}{\cal F}_{\nu \bar \mu}^b(x,\bar x)-{\cal D}^{ab\bar \nu}{\cal F}_{\bar \nu \bar \mu}^b(x,\bar x)+g_{d}f^{abc}{\cal A}^*_{c\bar \mu}(x,\bar x){\cal C}^b(x,\bar x)\, .
\end{align}
\end{subequations}

The maps given in Eqs.~(\ref{MF1},\ref{MF2}) accommodate representations of ${\rm SO}(1,3+n)$ into representations of the groups ${\rm SO}(1,3)$ and ${\rm SO}(n)$. However, to move completely from the extended symmetry to the standard one, we need to remove any dependence on the $\bar x$ coordinates from the theory. This step is nontrivial because, in the original theory, these coordinates play the role of labels that count degrees of freedom. In other words, we need to define another canonical map that allows us to hide any manifest dynamical role of the inhomogeneous ${\rm ISO}(n)$ subgroup. To do this, we will assign to each point $\bar x$ of the compact manifold that labels a field $\Phi^A(x,\bar x)$ or antifield $\Phi^*_{A}(x,\bar x)$ ($A=a\mu, \, a\bar \mu, \, a$) a linear combination of fields with well-defined transformation properties under the low energy groups ${\rm ISO}(1,3)\times {\rm SU}(N,{\cal M}^4)$. As shown in Ref.~\cite{OPRM}, this canonical mapping requires the specification of a compactification scheme and certain criteria to make contact with the standard theory. Following Ref.~\cite{OPRM}, the fields and antifields are expanded in terms of a basis of orthogonal functions $\{ f^{(\underline{0})}_E, f^{(\underline{k})}_E(\bar k \cdot \bar x), f^{(\underline{k})}_O(\bar k\cdot\bar x) \}$, with $\bar k\cdot \bar x=k_{\bar \mu} x_{\bar \mu}$, which are generated by the Casimir invariant $\bar{P}^2=P_{\bar \mu} P_{\bar \mu}$ of the inhomogeneous group $ISO(n)$. This set of eigenfunctions has definite parity under $\bar x\to -\bar x$, as they are multidimensional sines and cosines. Adopting the physical criterion that all fields with a standard counterpart are even, while those without a standard counterpart are odd, the set of fields $\{{\cal A}^{a\mu}, {\cal C}^a, \bar {\cal C}^a, {\cal B}^a \}$ is expanded in an even Fourier series, while only the fields ${\cal A}^{a \bar \mu}$ are mapped by an odd Fourier series~\cite{OPRM}, with identical expressions for the corresponding antifields. These requirements are satisfied by the compact manifold constructed from the direct product of $n$ orbifolds $S^1/\mathbb{Z}_2$. Periodicity, expressed as $\bar{x}_i = \bar{x}_i + 2\pi R$, arises from the geometry of the circle of radius $R$, while the $\mathbb{Z}_{2}$ cyclic group action identifies each point $\bar{x}_{i}$ with its antipode $-\bar{x}_{i}$ to define the parity. On this manifold, Dirichlet and Neumann boundary conditions lead to a discrete spectrum for $P_{\bar \mu }$ and $\bar{P}^{2}$, where the eigenvalues of the Casimir invariant $\bar{P}^{2}$ correspond to the masses of the KK excitations.\\

The gauge parameters also have a standard counterpart, therefore, they are expanded in an even Fourier series,
\begin{equation}
\alpha^a(x,\bar x)=f^{(\underline{0})}_E \alpha^{(\underline{0}) a}(x)+\sum_{(\underline{m})}f^{(\underline{m})}_E(\bar x) \alpha^{(\underline{m})a }(x)\, .
\end{equation}
As discussed in Ref.~\cite{OPRM}, the $\alpha^{(\underline{0}) a}(x)$ parameters define the standard gauge transformations (SGT), that is, they are the parameters of the standard gauge group ${\rm SU}(N,{\cal M}^4)$. On the other hand, the infinite set of parameters $\alpha^{(\underline{m})a }(x)$ define also gauge transformations, which, to distinguish them from the usual ones, we have called non-standard gauge transformations (NSGT). The NSGTs have their analogues in theories with spontaneous symmetry breaking  (see Ref. \cite{OPRM} ). Associated with each of the gauge fields $A^{(\underline{m})a}(x)$ there is a gauge parameter $\alpha^{(\underline{m})a}$.\\

By applying the Fourier mapping to the expressions (\ref{LG1},\ref{LG2}) and setting $\alpha^{(\underline{m})a}(x)=0$ in the resulting expressions, we obtain the way in which the various fields of classical theory are transformed under the standard gauge group ${\rm SU}(N,{\cal M}^4)$
\begin{subequations}
\begin{align}
\label{SGT0}
\delta_s A^{(\underline{0})a}_\mu (x)&={\cal D}^{(\underline{0})ab}_\mu \alpha^{(\underline{0})b}(x) \, , \\
\label{SGTm}
\delta_s A^{(\underline{m})a}_\mu (x)&=gf^{abc} A^{(\underline{m})b}_\mu (x) \alpha^{(\underline{0})c}(x)\, ,\\
\label{SGTSm}
\delta_s A^{(\underline{m})a}_{\bar \mu} (x)&=gf^{abc} A^{(\underline{m})b}_{\bar \mu} (x) \alpha^{(\underline{0})c}(x) \, ,
\end{align}
\end{subequations}
 where ${\cal D}^{(\underline{0})ab}_\mu=\delta^{ab}\partial^\mu-gf^{abc}A^{(\underline{0})c\mu}$ is the covariant derivative in the adjoint representation of the gauge group.  The above expressions show us that $A^{(\underline{0})a}_\mu (x)$ transform as gauge field, whereas $A^{(\underline{m})a}_\mu (x)$ and $A^{(\underline{m})a}_{\bar \mu} (x)$ transform as matter fields (in the adjoint representation) under this group. Thus, from the perspective of the standard group ${\rm SU}(N,{\cal M}^4)$, the KK excitations $A^{(\underline{m})a\mu}(x)$ can be endowed with mass, as occurs in theories with SSB. It is important to note that in obtaining the above results we have identified the dimensionless coupling constant $g$ of the standard gauge group ${\rm SU}(N,{\cal M}^4)$ with the dimensionful constant $g_d$ of the extended group ${\rm SU}(N,{\cal M}^d)$ through the following relation $g= f^{(\underline{0})}_E  g_d=\frac{g_d}{(2\pi R)^{\frac{n}{2}}}$. This association is crucial to make contact with standard theory. On the other hand, using the inverse of the Fourier maps, it is easy to show that the antibracket is preserved,$(\Phi^{(\underline{m})A}, \Phi^{*}_{(\underline{n})B})=\delta^{(\underline{m})}_{(\underline{n})}\delta^A_B$, thus showing that the maps that allow us to go from the extended symmetry ${\rm ISO}(1,3+n)\times {\rm SU}(N,{\cal M}^{d})$ to the standard symmetry $ {\rm ISO}(1,3)\times {\rm SU}(N,{\cal M}^{4})$ are canonical in the sense of the field-antifield formalism.\\

Implementing the maps given by the Fourier series in the proper solution, Eqs.~(\ref{PSF}) and (\ref{A2}), and integrating over the extra coordinates, we obtain
\begin{eqnarray}
\label{PSE}
S[\Phi,\Phi^*]&=&\int d^4x\, \Biggl \{ {\cal L}^{\rm YM}_{\rm eff}+A^*_{(\underline{0})a\mu}\left({\cal D}^{(\underline{0})ab\mu}C^{(\underline{0})b}-gf^{abc}\sum_{(\underline{m})}A^{(\underline{m})c\mu}C^{(\underline{m})b}  \right)\nonumber \\
&&+\sum_{(\underline{m})}A^*_{(\underline{m}) a\mu}\left(\sum_{(\underline{r})}{\cal D}^{(\underline{mr})ab\mu}C^{(\underline{r})b}-gf^{abc}A^{(\underline{m})c\mu}C^{(\underline{0})b}\right)\nonumber \\
&&+ \sum_{(\underline{m})} A^*_{(\underline{m})a\bar \mu}\left(\sum_{(\underline{r})} {\cal D}^{(\underline{mr})ab}_{\bar \mu}C^{(\underline{r})b}+gf^{abc}A^{(\underline{m}) c}_{\bar \mu}C^{(\underline{0})b} \right)\nonumber \\
&&+ \frac{1}{2}gf^{abc}\Biggl [C^*_{(\underline{0})c}\left(C^{(\underline{0})b}C^{(\underline{0})a}+\sum_{(\underline{m})}C^{(\underline{m})b}C^{(\underline{m})a} \right) \nonumber \\
&& +\sum_{(\underline{m})}C^*_{(\underline{m})c}\left(C^{(\underline{0})b}C^{(\underline{m})a}+C^{(\underline{m})b}C^{(\underline{0})a}+
\sum_{(\underline{nr})}\Delta_{(\underline{mnr})}C^{(\underline{n})b}C^{(\underline{r})a} \right)\Biggr]\nonumber \\
&&+\bar{C}^*_{(\underline{0})a}B^{(\underline{0})a}+\sum_{(\underline{m})}\bar{C}^*_{(\underline{m})a}B^{(\underline{m})a} \Biggr \}\, .
\end{eqnarray}
where ${\cal L}^{\rm YM}_{\rm eff}$ is the original classical Lagrangian arising from Eq.({\ref{A2}), whose gauge and Lorentz structures, along with notation and conventions,  have been discussed at length in Ref.~\cite{OPRM}. Here, we briefly discuss its essential pieces. This Lagrangian can be written as follows:
\begin{equation}
\label{EL}
 {\cal L}^{\rm YM}_{\rm eff}= {\cal L}^{\rm YM}_{\textrm{v-v}}+ {\cal L}^{\rm YM}_{\textrm{v-s}}+ {\cal L}^{\rm YM}_{\textrm{s-s}}\, .
\end{equation}
The Lagrangians $ {\cal L}^{\rm YM}_{\textrm{v-s}}$ and $ {\cal L}^{\rm YM}_{\textrm{s-s}}$ correspond to a scalar kinetic sector and to a scalar potential, respectively. The masses for the gauge, $A^{(\underline{m})a}_\mu(x)$, and scalar, $A^{(\underline{m})a}_{\bar \mu}(x)$, fields arise from the former and latter  Lagrangians, respectively. The Lagrangian $ {\cal L}^{\rm YM}_{\textrm{v-v}}$ contains the Yang-Mills term associated with the standard ${\rm SU}(N,{\cal M}^4)$ group plus terms involving interactions between gauge and matter fields. On the other hand, from the structure of the proper solution in Eq.~(\ref{PSE}), we can see that the terms involving only zero modes correspond to the standard result for the ${\rm SU}(N,{\cal M}^4)$ group.

\subsection{Mass spectrum}
\label{MS}
From the quadratic part in the Lagrangian $ {\cal L}^{\rm YM}_{\textrm{s-s}}$, massive scalars and massless scalars (pseudo--Goldstone bosons) arise. In the general case, the scalar fields $ A^{(\underline{k})a}_{\bar{\mu}} $ show themselves in bilinear forms, so that, after proper diagonalizations, one recognizes a pseudo-Goldstone boson, which we denote by $ A^{(\underline{k})a}_{\rm G} $, and $n-1$ massive scalar fields, represented here by $ A^{(\underline{k})a}_{\bar{n}} $, with $ \bar{n}=1,2,\ldots,n-1 $. The mass term can be written as follows~\cite{OPRM}:
\begin{equation}
\label{ELSS1}
{\cal L}^{\rm YM}_{\textrm{(s-s)}Mass}=-\sum_{(\underline{k})} \frac{1}{2}A^{(\underline{k})a}_{\bar \mu}\mathfrak{M}_{\bar \mu \bar \nu }^{(\underline{k})}A^{(\underline{k})a}_{\bar \nu}
\end{equation}
where, for each possible value of  $ (\underline{k}) $, the corresponding $ n\times n $ symmetric mass matrix is
\begin{equation}
\label{MM}
\mathfrak{M}^{(\underline{k})}_{\bar{\mu}\bar{\nu}}=m_{(\underline{k})}^{2}\delta_{\bar{\mu}\bar{\nu}}-p_{\bar{\mu}}^{(\underline k)}  p_{\bar{\nu}}^{(\underline k)}\, .
\end{equation}
In this expression, the discrete momentum $p_{\bar \mu}$ arises from $\partial_{\bar \mu}f_{E,O}^{(\underline{k})}(\bar p \cdot \bar x)=\mp p^{(\underline{k})}_{\bar \mu}f_{O,E}^{(\underline{k})}(\bar p \cdot \bar x)$ . Note that $m_{(\underline{k})}=\sqrt{p^{(\underline{k})}_{\bar \mu}p^{(\underline{k})}_{\bar \mu}}$. The above matrix, whose mathematical structure is directly dictated by gauge invariance, naturally leads to the existence of a massless scalar field. To see this, we take the trace in Eq. (\ref{MM}) to obtain
\begin{equation}
\delta^{\bar \mu  \bar \nu}\mathfrak{M}^{(\underline{k})}_{\bar{\mu}\bar{\nu}}=(n-1)m^2_{(\underline{k})}\, ,
\end{equation}
which shows that there exist $(n-1)$ mass-degenerate physical scalar fields and a massless scalar field, which can be identified with a pseudo-Goldstone boson. This matrix can
be diagonalized through an orthogonal transformation given by a matrix $\mathcal{R}^{(\underline{k})}$~\cite{OPRM}.\\

In terms of the mass eigenstates, the scalar sector of the classical theory becomes
\begin{equation}
{\cal L}^{\rm YM}_{\textrm{s-s}}=-\sum_{(\underline{m})}\Bigg\{ \frac{1}{2}m^2_{(\underline{m})}A^{(\underline{m})a}_{\bar n}A^{(\underline{m})a}_{\bar n} +\cdots \Bigg\}\, .
\end{equation}
From this Lagrangian we immediately recognize the massive fields $ A^{(\underline{m})a}_{\bar{n}} $, with mass $ m_{(\underline{m})} $. The ellipsis represent various trilinear- and quartic-interaction terms among them. It is worth emphasizing that all the mass terms are generated by the curvature components ${\cal F}^{(\underline{m})a}_{{\bar \mu}\nu}(x)$ and ${\cal F}^{(\underline{m})a}_{{\bar \mu}{\bar \nu}}(x)$, which in turn come from the extra-dimensional curvature ${\cal F}^a_{MN}(x,\bar x)$, whose precise structure is fixed by gauge invariance. So all masses originate in gauge symmetry and, in that sense, they can be properly called ``gauge masses''. This is the reason of the mass degeneration. As far as the massless scalar $ A^{(\underline{m})a}_{\rm G} $ fields are concerned, they play the role of pseudo-Goldstone bosons, as they can be removed from the theory through some sort of unitary gauge~\cite{OPRM}. The degrees of freedom that they represent appear as the longitudinal polarization states of the vector KK excitations $ A^{(\underline{m})a}_{\mu}$~\cite{OPRM}. Note that the ${\cal L}^{\rm YM}_{\textrm{s-s}}$ Lagrangian plays the role of a Higgs potential, as it allows us to determine the presence of pseudo-Goldstone bosons and physical scalars.\\

The Lagrangian ${\cal L}^{\rm YM}_{\textrm{v-s}}$ is also affected by the orthogonal transformation given by the matrix $\mathcal{R}^{(\underline{k})}$. One has,
\begin{eqnarray}
\label{ELVS2}
{\cal L}^{\rm YM}_{\textrm{v-s}}&=&\sum_{(\underline{m})}\Bigg\{\frac{1}{2}\left({\cal D}^{(\underline{0})ab}_\mu A^{(\underline{m})b}_{\bar n}\right)\left({\cal D}^{(\underline{0})ac\mu} A^{(\underline{m})c}_{\bar n}\right)+\frac{1}{2}\left({\cal D}^{(\underline{0})ab}_\mu A^{(\underline{m})b}_{\rm G}\right)\left({\cal D}^{(\underline{0})ac\mu} A^{(\underline{m})c}_{\rm G}\right)\nonumber \\
&&+m_{(\underline{m})}A^{(\underline{m})a\mu}\left({\cal D}^{(\underline{0})ab}_\mu A^{(\underline{m})b}_{\rm G}\right)+\frac{1}{2}m^2_{(\underline{m})} A^{(\underline{m})a}_\mu A^{(\underline{m})a\mu}+\cdots\, ,
\Bigg\}\, .
\end{eqnarray}
where the ellipsis represent trilinear and quartic interactions between scalar and vector fields~\cite{OPRM}. Note the presence of the mass term for the gauge KK excitations. Also from the term proportional to the $m_{(\underline{m})}$ scale in the above expression, one can identify the presence of bilinear interactions between the vector KK excitations $A^{(\underline{m})a}_\mu$ and their associated pseudo-Goldstone bosons $A^{(\underline{m})a}_{\rm G}$. This class of interactions is typical of Higgs kinetic terms. We will show below how to remove this term from the theory using an appropriate gauge-fixing  procedure.

\subsection{The gauge-fixed action $S_\Psi$}
\label{GFPSI}
Deriving a gauge-invariant one-loop effective action requires the implementation of a gauge-fixing procedure for the fields  $A^{(\underline{0})a\mu}$  and $A^{(\underline{k})a\mu}$ that preserves covariance under the $SU(N,{\cal M}^4)$ group. One way to achieve this is by using the BFM~\cite{BFM}, which allows gauge invariance to be preserved throughout the calculation. The BFM consists of decomposing each field $\Phi^A$ into a classical part, $\bar{\Phi}^A$, and a quantum part, $\phi^A$ (that is, $\Phi^A$ is replaced by $\bar{\Phi}^A+\phi^A $). While a complete study of the theory~\cite{DA,BBHSS} requires all fields to be decomposed, we focus here only on the one-loop effects of KK excitations. Accordingly, classical parts for the KK excitations are omitted. Therefore, only the standard fields $A^{(\underline{0})a\mu}$ will decompose into their classical and quantum parts, that is, we will implement the change $A^{(\underline{0})a\mu}=\bar{A}^{(\underline{0})a\mu}+Q^{(\underline{0})a \mu}$ everywhere, being $Q^{(\underline{0})a\mu}$ the quantum part.\\

With this decomposition, the SGT given in Eq.~(\ref{SGT0}) leads to the following relations:
\begin{subequations}
\begin{align}
\label{SGT0BF}
\delta_s \bar{A}^{(\underline{0})a}_\mu&= \bar{{\cal D}}^{(\underline{0})ab}_\mu \alpha^{(\underline{0})b}\, , \\
\label{SGT0QF}
\delta_s Q^{(\underline{0})a}_\mu&= gf^{abc}Q^{(\underline{0})b}_\mu \alpha^{(\underline{0})c}\, ,
\end{align}
\end{subequations}
where
\begin{equation}
\label{CDBF}
\bar{{\cal D}}^{(\underline{0})ab}_\mu=\delta^{ab}\partial_\mu-gf^{abc}\bar{A}^{(\underline{0})c}_\mu\, .
\end{equation}
From the above expressions, we can see that the background field transforms as a gauge field, while the quantum field transforms as a matter field in the adjoint representation. Note that $Q^{(\underline{0})a}_\mu$ and all KK excitations transform in the same representation of the standard group ${\rm SU}(N,{\cal M}^4)$.\\

In the BFM, the curvatures $F^{(\underline{0})a}_{\mu \nu}$ that appear in the Lagrangian ${\cal L}^{\rm YM}_{\rm eff}$ are also divided into a classical, $\bar{F}^{(\underline{0})a}_{\mu \nu}$, and a quantum, $Q^{(\underline{0})a}_{\mu \nu}$, parts, as follows:
\begin{equation}
F^{(\underline{0})a}_{\mu \nu}=\bar{F}^{(\underline{0})a}_{\mu \nu}+Q^{(\underline{0})a}_{\mu \nu}\, ,
\end{equation}
where
\begin{equation}
Q^{(\underline{0})a}_{\mu \nu}=\bar{{\cal D}}^{(\underline{0})ab}_\mu Q^{(\underline{0})b}_\nu-\bar{{\cal D}}^{(\underline{0})ab}_\nu Q^{(\underline{0})b}_\mu+gf^{abc}Q^{(\underline{0})b}_\mu Q^{(\underline{0})c}_\nu\, .
\end{equation}
The modified classical Lagrangian ${\cal L}^{\rm YM}_{\rm eff}={\cal L}^{\rm YM}_{\rm eff}(\bar{A}+Q)$ is invariant under the gauge transformations given by Eqs.~(\ref{SGT0BF}), (\ref{SGT0QF}), (\ref{SGTm}), and (\ref{SGTSm}).\\

The change $A^{(\underline{0})a\mu}=\bar{A}^{(\underline{0})a\mu}+Q^{(\underline{0})a\mu}$ leads to the following changes in the extended action (\ref{PSE}):
\begin{eqnarray}
\label{PSEBF}
S[\Phi,\Phi^*]&=&\int d^4x\, \Biggl \{ {\cal L}^{\rm YM}_{\rm eff}\left(\bar{A}^{(\underline{0})}+Q^{(\underline{0})}\right)\nonumber \\ &&+Q^*_{(\underline{0})a\mu}\left({\cal D}^{(\underline{0})ab\mu}\left(\bar{A}^{(\underline{0})}+Q^{(\underline{0})}\right)C^{(\underline{0})b}-gf^{abc}\sum_{(\underline{m})}A^{(\underline{m})c\mu}C^{(\underline{m})b}  \right)\nonumber \\
&&+\sum_{(\underline{m})}A^*_{(\underline{m}) a\mu}\left(\sum_{(\underline{r})}{\cal D}^{(\underline{mr})ab\mu}\left(\bar{A}^{(\underline{0})}+Q^{(\underline{0})}\right)C^{(\underline{r})b}-gf^{abc}A^{(\underline{m})c\mu}C^{(\underline{0})b}\right)\nonumber \\
&&+ \sum_{(\underline{m})} A^*_{(\underline{m})a\bar \mu}\left(\sum_{(\underline{r})} {\cal D}^{(\underline{mr})ab}_{\bar \mu}C^{(\underline{r})b}+gf^{abc}A^{(\underline{m}) c}_{\bar \mu}C^{(\underline{0})b} \right)\nonumber \\
&&+ \frac{1}{2}gf^{abc}\Biggl [C^*_{(\underline{0})c}\left(C^{(\underline{0})b}C^{(\underline{0})a}+\sum_{(\underline{m})}C^{(\underline{m})b}C^{(\underline{m})a} \right) \nonumber \\
&& +\sum_{(\underline{m})}C^*_{(\underline{m})c}\left(C^{(\underline{0})b}C^{(\underline{m})a}+C^{(\underline{m})b}C^{(\underline{0})a}+
\sum_{(\underline{nr})}\Delta_{(\underline{mnr})}C^{(\underline{n})b}C^{(\underline{r})a} \right)\Biggr]\nonumber \\
&&+\bar{C}^*_{(\underline{0})a}B^{(\underline{0})a}+\sum_{(\underline{m})}\bar{C}^*_{(\underline{m})a}B^{(\underline{m})a} \Biggr \}\, ,
\end{eqnarray}
where
\begin{equation}
{\cal D}^{(\underline{0})ab\mu}\left(\bar{A}^{(\underline{0})}+Q^{(\underline{0})}\right)=\bar{{\cal D}}^{(\underline{0})ab\mu}-gf^{abc}Q^{(\underline{0})c\mu} \, .
\end{equation}

On the other hand, after using the Fourier maps, the fermion functional (\ref{FerFun}) becomes
\begin{equation}
\label{FerFun2}
\Psi=\int d^4x\left[\bar{C}^{(\underline{0})a}\left(f^{(\underline{0})a}+\frac{\xi}{2}B^{(\underline{0})a}\right)+
\sum_{(\underline{m})}\bar{C}^{(\underline{m})a}\left(f^{(\underline{m})a}+\frac{\xi}{2}B^{(\underline{m})a}\right)\right]\, ,
\end{equation}
where an even Fourier map has been assumed for $f^a(x,\bar x)$, since it has a standard counterpart.\\

Now we are ready to introduce the gauge-fixing functions. For the standard quantum gauge fields $Q^{(\underline{0})a}_\mu$, we introduce a ${\rm SU}(N,{\cal M}^4)$-covariant gauge-fixing procedure through the following functions:
\begin{equation}
\label{GFf0}
f^{(\underline{0})a}=\bar{{\cal D}}^{(\underline{0})ab}_\mu Q^{(\underline{0})b\mu}\, ,
\end{equation}
which transforms in the adjoint representation of the group.\\

On the other hand, to define the propagators of the gauge bosons $A^{(\underline{m})a\mu}(x)$ and their associated pseudo-Goldstone bosons $A^{(\underline{ m})a}_{\rm G}(x)$, we need to eliminate the bilinear mixing between these fields, which appears in the Lagrangian ${\cal L}_{\textrm{v-s}}$, given in Eq.(\ref{ELVS2}). Such term is given by:
\begin{equation}
\label{mix}
{\cal L}^{\rm mix}_{\textrm{v-s}}=\sum_{(\underline{m})}m_{(\underline{m})}A^{(\underline{m})a\mu}\left(\bar{{\cal D}}^{(\underline{0})ab}_\mu A^{(\underline{m})b}_{\rm G}\right)\, .
\end{equation}
It is desirable to define a gauge-fixing procedure that, in addition to removing the bilinear terms $A^{(\underline{m})a \mu}A^{(\underline{m}) a}_{\rm G}$, allows us to remove also the vertices $A^{(\underline{m})a \mu}A^{(\underline{m}) a}_{\rm G} \bar{A}^{(\underline{0})a \mu}$. To this end, we introduce a ${\rm SU}(N,{\cal M}^4)$-covariant gauge-fixing procedure through the following gauge functions:
\begin{equation}
\label{GFfm}
f^{(\underline{m})a}=\bar{{\cal D}}^{(\underline{0})ab}_\mu A^{(\underline{m})b\mu}-\xi p^{(\underline{m})}_{\bar \mu}A^{(\underline{m})a}_{\bar \mu}\, .
\end{equation}
 Note that these gauge-fixing functions transform under the adjoint representation of the ${\rm SU}(N,{\cal M}^4)$ gauge group. This gauge-fixing  procedure exactly cancels the non-physical term ${\cal L}^{\rm mix}_{\textrm{v-s}}$. Actually, the gauge-fixing functions (\ref{GFfm}) do not depend on the physical scalar fields $A^{(\underline{m})a}_{\bar n}$, but only on the pseudo-Goldstone bosons. In fact, after using some properties of the orthogonal transformation implemented in the scalar sector~\cite{OPRM}, Eq.(\ref{GFfm}) becomes
\begin{equation}
\label{GFfm2}
f^{(\underline{m})a}=\bar{{\cal D}}^{(\underline{0})ab}_\mu A^{(\underline{m})b\mu}-\xi m_{(\underline{m})}A^{(\underline{m})a}_{G}\, ,
\end{equation}

Having defined the fermion $\Psi$ functional, we are ready to eliminate the antifields from the extended action $S$. We start with the usual fields. For the usual fields we have
\begin{subequations}
\begin{align}
\label{AF01}
Q^*_{(\underline{0})a\mu}&=-\bar{{\cal D}}^{(\underline{0})ab}_\mu \bar{C}^{(\underline{0})a}\, , \ \ \ C^*_{(\underline{0})a}=0\,  ,\\
\bar{C}^*_{(\underline{0})a}&=f^{(\underline{0})a}+\frac{\xi}{2}B^{(\underline{0})a}\, , \, \, B^*_{(\underline{0})a}=\frac{\xi}{2}\bar{C}^{(\underline{0})a}\, .
\end{align}
\end{subequations}

As far as the antifields of the KK excitations are concerned, we have
\begin{subequations}
\begin{align}
A^*_{(\underline{m})a\mu}&=-\bar{{\cal D}}^{(\underline{0})ab}_\mu \bar{C}^{(\underline{m})b}\, , \, \, A^*_{(\underline{m})a\bar{\mu}}=\xi p^{(\underline{m})}_{\bar \mu}\bar{C}^{(\underline{m})a}\, ,\\
C^*_{(\underline{m})a}&=0\, ,\, \, \bar{C}^*_{(\underline{m})a}=f^{(\underline{m})a}+\frac{\xi}{2}B^{(\underline{m})a}\, , \, \, B^*_{(\underline{m})a}=\frac{\xi}{2}\bar{C}^{(\underline{m})a}\, .
\end{align}
\end{subequations}
The replacement of the antifields given by the above expressions in the extended action (\ref{PSEBF}) leads to the gauge-fixed action $S_\Psi=S[\Phi,\frac{\delta \Psi}{\delta \Phi}]$. Once this is done and the auxiliary fields have been eliminated using the equations of motion, we obtain
\begin{equation}
\label{GFA}
S_\Psi=\int d^4x\, \left[ {\cal L}^{\rm YM}_{\rm eff}\left(\bar{A}^{(\underline{0})}+Q^{(\underline{0})}\right)+{\cal L}_{GF}+{\cal L}_{\rm FPG}\right]\, ,
\end{equation}
where ${\cal L}_{GF}$ and ${\cal L}_{\rm FPG}$ are the gauge-fixing and Faddeev-Popov ghost Lagrangians, respectively, which are given by
\begin{equation}
\label{GF}
{\cal L}_{GF}=-\frac{1}{2\xi}f^{(\underline{0})a}f^{(\underline{0})a}-\frac{1}{2\xi}\sum_{(\underline{m})}f^{(\underline{m})a}f^{(\underline{m})a}\, ,
\end{equation}
\begin{eqnarray}
\label{FPG}
{\cal L}_{\rm FPG}&=&-\bar{{\cal D}}^{(\underline{0})ab}_\mu \bar{C}^{(\underline{0})b}\Biggl({\cal D}^{(\underline{0})ab\mu}\left(\bar{A}^{(\underline{0})}+Q^{(\underline{0})}\right)C^{(\underline{0})b}
-gf^{abc}\sum_{(\underline{m})}A^{(\underline{m})c\mu}C^{(\underline{m})b}  \Biggr)\nonumber \\
&&-\sum_{(\underline{m})}\bar{{\cal D}}^{(\underline{0})ab}_\mu \bar{C}^{(\underline{m})b}\left(\sum_{(\underline{r})}{\cal D}^{(\underline{mr})ab\mu}\left(\bar{A}^{(\underline{0})}+Q^{(\underline{0})}\right)C^{(\underline{r})b}-gf^{abc}A^{(\underline{m})c\mu}C^{(\underline{0})b}\right)\nonumber \\
&&+ \sum_{(\underline{m})}\Biggl(\xi m^2_{(\underline{m})} \bar{C}^{(\underline{m})a}C^{(\underline{m})a}+\xi g f^{abc}m_{(\underline{m})}A^{(\underline{m})c}_G\bar{C}^{(\underline{m})a}C^{(\underline{0})b}+\xi p^{(\underline{m})}_{\bar \mu}\sum_{(\underline{rs})}\Delta'_{(\underline{mrs})}
A^{(\underline{s})b}_{\bar \mu}\bar{C}^{(\underline{m})a}C^{(\underline{r})b}
\Biggr)\, .
\end{eqnarray}

As already commented, this gauge-fixing  procedure for the KK gauge fields exactly cancels the mixing term displayed in Eq.~(\ref{mix}),
\begin{equation}
{\cal L}^{\rm mix}_{\textrm{v-s}}-\frac{1}{2\xi}\sum_{(\underline{m})}f^{(\underline{m})a}f^{(\underline{m})a}=-\sum_{(\underline{m})}\Biggl[\frac{1}{2\xi} \left(\bar{{\cal D}}^{(\underline{0})ab}_\mu A^{(\underline{m})b\mu}\right)^2+\frac{1}{2}\xi m^2_{(\underline{m})}\left(A^{(\underline{m})a}_{\rm G}\right)^2\Biggr]\, .
\end{equation}
Note that, in addition to defining the propagators of the gauge fields $A^{(\underline{m})a}_\mu$, the first term of this expression introduces non-trivial modifications to the trilinear, $A^{( \underline {m})a}A^{(\underline{m})a} \bar{A}^{(\underline{0})a}$, and quartic, $A^{(\underline{m})a} A^ {(\underline{m})a} \bar{A}^{(\underline{0})a}\bar{A}^{(\underline{0})a}$, vertices that appear in the ${\cal L}^ {\rm YM }_{\textrm{v-v}}$ Lagrangian. The last term of this expression introduces a nonphysical mass for the pseudo-Goldstone bosons $A^{(\underline{m})a}_{\rm G}$. Everything works as in theories with SSB.

\section{Conclusions}
\label{C}In this paper, we investigated the structure of a pure Yang-Mills theory with $n$ flat extra dimensions, quantized via the Batalin-Vilkovisky (BV) or BRST formalism. By embedding the $(4+n)$-dimensional Yang-Mills theory based on ${\rm ISO}(1,3+n)\times {\rm SU}(N,{\cal M}^{4+n})$ within the field-antifield framework, we derived a real, bosonic, ghost-number-zero solution to the classical master equation $(S,S)=0$. To resolve the degeneracy of this solution and achieve covariant quantization, we executed a gauge-fixing procedure using a fermionic functional $\Psi$ that incorporates antighosts, auxiliary fields, and arbitrary gauge-fixing functions $f^a(x,\bar x)$.\\

To describe low-energy physical phenomena at the compactification scale $R^{-1}$, we utilized the concept of hidden symmetries alongside compactification to generate masses for the Kaluza-Klein (KK) degrees of freedom. The transition from the high-energy to the low-energy regime was achieved by breaking the original ${\rm ISO}(1,3+n)\times {\rm SU}(N,{\cal M}^{4+n})$ symmetries into ${\rm ISO}(1,3)\times {\rm SU}(N,{\cal M}^4)$ via two canonical transformations within the antibracket framework. Furthermore, we introduced a manifestly covariant gauge-fixing procedure that preserves the standard $SU(N,{\cal M}^{4})$ symmetry. Specifically, the standard contributions were fixed using the BFM, while the KK excitations were handled by exploiting their transformation properties as matter fields in the adjoint representation of the gauge group. With this robust, gauge-invariant, and manifestly covariant framework established, the theory is now fully poised to investigate the one-loop corrections induced by KK fluctuations on standard observables. Work is currently underway to analyze the one-loop ultraviolet structure of the theory and its renormalization within the context of effective field theories.

\section*{Acknowledgments}
We acknowledge financial support from SECIHTI (Secretar\'\i a de Ciencia, Humanidades, Tecnolog\'\i a e Innovaci\' on) and SNII (Sistema Nacional de Investigadoras e Investigadores) (M\' exico). We also acknowledge financial support from Vicerrector\'ia de Investigaci\'on y Estudios de Posgrado, Benem\'erita Universidad Aut\'onoma de Puebla, through Project No. 100525129-VIEP2026.

\end{document}